\documentclass[twocolumn,english,singlespace]{revtex4-1}
\usepackage[T1]{fontenc}
\usepackage[latin9]{inputenc}
\usepackage{color}
\usepackage{amstext}
\usepackage{amssymb}
\usepackage{graphicx}

\makeatletter

\@ifundefined{textcolor}{}
{%
 \definecolor{BLACK}{gray}{0}
 \definecolor{WHITE}{gray}{1}
 \definecolor{RED}{rgb}{1,0,0}
 \definecolor{GREEN}{rgb}{0,1,0}
 \definecolor{BLUE}{rgb}{0,0,1}
 \definecolor{CYAN}{cmyk}{1,0,0,0}
 \definecolor{MAGENTA}{cmyk}{0,1,0,0}
 \definecolor{YELLOW}{cmyk}{0,0,1,0}
}

\@ifundefined{definecolor}
 {\usepackage{color}}{}

\makeatother

\usepackage{babel}
\begin{document}

\title{Secure continuous variable teleportation and Einstein-Podolsky-Rosen
steering}

\author{Q. Y. He$^{1,2}$, L. Rosales-Zárate$^{2}$, G. Adesso$^{3}$ and
M. D. Reid$^{2}$}

\affiliation{$^{1}$State Key Laboratory of Mesoscopic Physics, School of Physics,
Peking University, Beijing 100871 China }

\email{qiongyihe@pku.edu.cn}

\affiliation{$^{\text{2}}$Centre for Quantum and Optical Science, Swinburne University
of Technology, Melbourne 3122 Australia}

\email{lrosaleszarate@swin.edu.au}

\email{mdreid@swin.edu.au}

\affiliation{$^{\text{3}}$School of Mathematical Sciences, University of Nottingham,
Nottingham NG7 2RD, United Kingdom }

\email{gerardo.adesso@nottingham.ac.uk}

\begin{abstract}
We investigate the resources needed for secure teleportation of coherent
states. We extend continuous variable teleportation to include quantum
tele-amplification protocols, that allow non-unity classical gains
and a pre-amplification or post-attenuation of the coherent state.
We show that, for arbitrary Gaussian protocols and a significant class
of Gaussian resources, two-way steering is required to achieve a teleportation
fidelity beyond the no-cloning threshold. This provides an operational
connection between Gaussian steerability and secure teleportation.
We present practical recipes suggesting that heralded noiseless pre-amplification
may enable high-fidelity heralded teleportation, using minimally entangled
yet steerable resources. 
\end{abstract}
\maketitle
Quantum teleportation (QT) is a process where Alice sends an unknown
quantum state to Bob at a different location by communicating only
classical information \cite{telebenentt}. QT has inspired much interest,
both as a fundamental challenge and as a tool for quantum information
processing \cite{bowtele,cvteleexpfuru,zhangtele,yone,vaid,bk,bou}.
To achieve QT, Alice and Bob share an Einstein-Podolsky-Rosen (EPR)
entangled state. Teleportation was first developed for the transfer
of qubit states, and was extended to continuous variable (CV) spectra
by Vaidman \cite{vaid} and Braunstein and Kimble (BK) \cite{bk}.
In the CV case, the entanglement shared between Alice and Bob is modeled
after the original EPR paradox where Alice and Bob share systems with
perfectly correlated positions and anti-correlated momenta \cite{EPR paradox,two-modeeprresource,eprcrit,ou}.
Gaussian states (defined as having a Gaussian Wigner function) \cite{gauss}
can then be useful as approximations of EPR resources \cite{bowtele,cvteleexpfuru}.

{What type of EPR entanglement is required for CV quantum teleportation
\cite{horoquestion}?} CV teleportation of a coherent state originally
focused on a subset of entangled resource states, where the entanglement
can be certified by the Tan-Duan criterion which treats Alice and
Bob symmetrically \cite{Tan,duan-1,buono}: 
\begin{equation}
\Delta_{ent}=\frac{1}{4}\{[\Delta(X_{A}-X_{B})]^{2}+[\Delta(P_{A}+P_{B})]^{2}\}<1.\label{eq:duan-1-1-1}
\end{equation}
Here $X_{A}$,$X_{B}$ and $P_{A},$$P_{B}$ are the positions and
momenta of Alice and Bob's systems and $\Delta X^{2}$ denotes the
variance of $X$ \cite{notevariance}. Once one allowed for local
operations at Alice and Bob's stations to optimise the protocol, it
became clear that all two-mode Gaussian entangled states could be
utilised for CV QT \cite{adessotele,mari} with fidelity $F$ exceeding
$1/2$, which is the standard benchmark for input coherent states
\cite{ham}.

These results however do not resolve a second fundamental question,
posed by Grosshans and Grangier (GG) \cite{Grangier clone}: What
type of entangled state is required, if Bob is to be sure there can
be no (non-degraded) copy of his transmitted (coherent) state in the
hands of a second receiver, Eve? This form of entanglement becomes
the vital resource for quantum information tasks where one needs \textit{secure}
teleportation (ST). An analysis based on optimal quantum cloning tells
us that, for coherent inputs, ST is achieved once the fidelity $F$
of CV teleportation exceeds $2/3$ \cite{Grangier clone,cerfclone}.

Here we solve such a longstanding question by proving that secure
CV teleportation requires a stronger form of entanglement exhibiting
\emph{EPR steering} \cite{hw-steering,ericsteer,saunexp}. EPR steering
refers to the correlations of the original 1935 EPR paradox \cite{EPR paradox},
where one observer appears to adjust (``steer'') the state of the
other by local measurements. A useful criterion to certify the EPR
paradox and a steering of $B$ by $A$ in a bipartite state is \cite{eprcrit,rmp-1}
\begin{equation}
E{}_{B|A}(\mathbf{g})=\Delta(X_{B}-g_{x}X_{A})\Delta(P_{B}+g_{p}P_{A})<1.\label{eq:eprsteercrit}
\end{equation}
Here $\mathbf{g}=(g_{x},g_{p})$ where $g_{x}$, $g_{p}$ are real
constants, usually chosen so that $E_{B|A}(\mathbf{g})$ is the minimum
possible value, denoted $E_{B|A}$. Then, $B$ is steerable by $A$
if $E_{B|A}<1$. This condition is necessary and sufficient for two-mode
Gaussian states \cite{hw-steering,adesso2}, and a quantifier of Gaussian
EPR steering can be defined as a decreasing function of $E_{B|A}$
\cite{adesso}. Clearly, for steering, as in the original formulation
of the EPR paradox, Alice and Bob are \emph{not} equivalent: A system
can be steerable in one way but not the other \cite{adesso,qhediscord,two way steering,Bogdan_CVonesided,piani}.
The role of steering in teleportation was first explored by GG, who
noticed that, for the original BK protocol, resources displaying EPR
paradox correlations are needed to achieve ST \cite{Grangier clone}.
However, the generality of such a conclusion remained unclear.

Motivated by the recent progress in EPR steering characterization
and quantification \cite{hw-steering,ericsteer,rmp-1,two way steering,saunexp,navascues,Bogdan_CVonesided,piani,qhediscord,adesso,adesso2,adesso3,steer telequbit},
we revisit CV teleportation by considering asymmetric Gaussian resources
and the whole class of protocols including those allowing an arbitrary
classical gain and local pre-amplification or post-attenuation of
the coherent state. For a significant class of practical resources,
simple arguments reveal that ST requires a one-way ($A$ by $B$)
steerable resource. This leads us to formulate and prove the generalisation
of the GG result: Any Gaussian resource which is useful for high fidelity
ST $(F>2/3)$ via an optimal protocol is necessarily two-way steerable.
We further show how for particular protocols and resource states this
condition becomes also sufficient.

We clarify the trade-off between required entanglement and achievable
ST fidelity. While two-way steerability requires exceeding a threshold
in entanglement, the latter becomes low for states of sufficient purity
\cite{adesso}. We propose that such states, if combined with a heralded
noiseless pre-amplification of the coherent state \cite{heraldednoiseless},
can nonetheless be useful for realising high fidelity ST. In this
way, our work may contribute to the practical problem of how the fidelity
of CV teleportation can be improved without increasing the entanglement
(and hence energy requirements) of the EPR resource \cite{recent high fidelity}.

\textbf{\emph{Quantum tele-amplification: }}We begin by considering
the generalisation of the BK protocol \cite{bk}, which incorporates
arbitrary classical gains (Fig. 1). As with conventional CV teleportation,
Alice and Bob share an EPR entangled state, often modeled by the two-mode
squeezed state (TMSS) $|\psi\rangle=(1-x^{2})^{1/2}\sum_{n=0}^{\infty}x^{n}|n\rangle_{A}|n\rangle_{B}$
\cite{two-modeeprresource,eprcrit,notevariance}. Here \textcolor{black}{$x=\tanh(r)$,
where $r$} is the squeezing parameter that determines the amount
of entanglement shared between Alice and Bob; the limit of maximal
EPR entanglement is reached as $r\rightarrow\infty$. Realistic conditions
(e.g. losses) mean that the shared EPR resource is best described
as a two-mode Gaussian state \cite{buono,gauss,rmp-1,mari,adessotele}.
A field $V$ (with amplitudes $X_{V}$ and $P_{V}$) is prepared by
a third party, Victor, in the coherent state $|\psi_{in}\rangle=|\alpha\rangle$
that is to be teleported to Bob. Alice performs a local Bell measurement
of the combined quadratures $X_{V}-X_{A}$ and $P_{V}+P_{A}$, to
give outcomes $m_{x}$ and $m_{p}$ respectively. The final stage
of the teleportation is the displacement by Bob of the amplitudes
of his EPR field by an amount given by Alice's readout values $m_{x}$,
$m_{p}$, that are transmitted to him from Alice using classical communication.

\begin{figure}
\begin{centering}
\includegraphics[width=0.7\columnwidth]{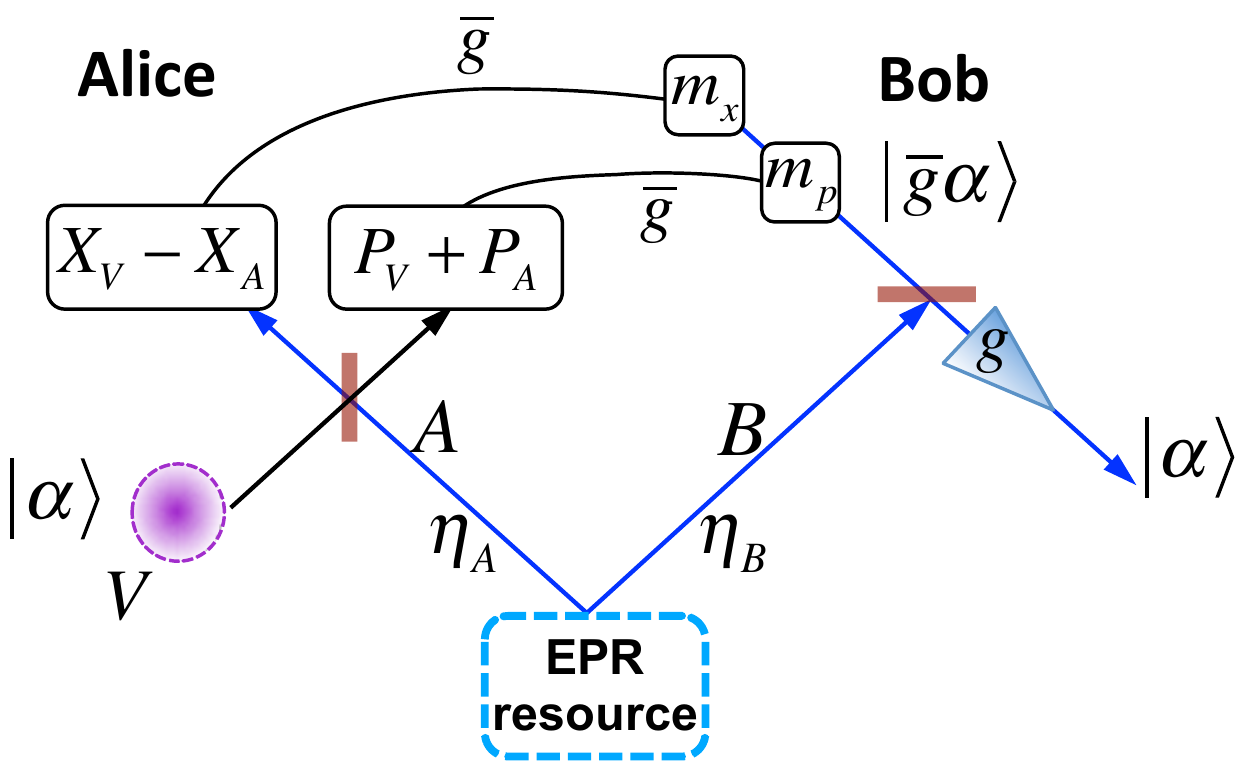} 
\par\end{centering}

\protect\protect\protect\caption{\textcolor{black}{\emph{Quantum tele-amplification. }}\textcolor{black}{A
coherent state $|\alpha\rangle$ is teleported from Alice to Bob using
EPR entanglement.} The fidelity is optimised for a given resource
by adjusting the classical gain $\bar{g}$. \label{fig:BK}To teleport
the original state, Bob may post-attenuate the state using $g=1/\bar{g}$
or Alice may pre-amplify the coherent state.}
\end{figure}

While the BK protocol takes $\bar{g}=1$, we allow for non-unity classical
gain factors $\bar{g}_{x}$, $\bar{g}_{p}$ in the two classical channels.
For simplicity, we consider equal gains, $\bar{g}_{x}=\bar{g}_{p}=\bar{g}$.
This means that Bob's displacement is amplified/ deamplified to $\bar{g}m_{x}$
and $\bar{g}m_{p}$. After feedback, Bob's field amplitudes are given
by $X_{B}^{f}=\bar{g}X_{V}+(X_{B}-\bar{g}X_{A}),\ P_{B}^{f}=\bar{g}P_{V}+(P_{B}+\bar{g}P_{A}).$

\textcolor{black}{Initially, we evaluate the fidelity for the protocol
}$|\alpha\rangle\rightarrow|\bar{g}\alpha\rangle$\textcolor{black}{,
called} ``quantum tele-amplification'' when $\bar{g}>1$ \cite{tele-amp,qhediscord}\textcolor{black}{.
Then, the desired teleported state is $|\beta_{tele}\rangle=|\bar{g}\alpha\rangle$.
}The fidelity, defined as $F=\langle\beta_{tele}|\rho_{out}|\beta_{tele}\rangle$
where $\rho_{out}$ is the density operator of the output state at
Bob's location, \textcolor{black}{is calculated using standard techniques
}\cite{bk,fidtele,zhangtele}\textcolor{black}{. }The result is $F=\frac{2}{\sigma_{Q}}\exp\left[-\frac{2}{\sigma_{Q}}\bigl|\beta_{out}-\beta_{tele}\bigr|^{2}\right]$\textcolor{black}{{}
\cite{zhangtele} where}\textcolor{red}{{} }$\sigma_{Q}=\sqrt{(1+\sigma_{X})(1+\sigma_{P})}$
and $\beta_{out}=x_{m}+ip_{m}$. Here \textcolor{black}{$x_{m}$,
$p_{m}$ and }$\sigma_{X}$, $\sigma_{P}$\textcolor{black}{{} are
the means and variances}\textcolor{green}{{} }of the quadratures
\textcolor{black}{$X_{B}^{f}$, $P_{B}^{f}$ }of Bob's output field.
We find\textcolor{black}{{} }$\sigma_{X}=\bar{g}^{2}\sigma_{X,in}+[\Delta(X_{B}-\bar{g}X_{A})]^{2}$\textcolor{black}{,
$\sigma_{P}=\bar{g}^{2}\sigma_{P,in}+[\Delta(P_{B}+\bar{g}P_{A})]^{2}$,
where $\sigma_{X/P,in}$ }is the variance of $X$/ $P$ for the input
state ($|\alpha\rangle$) at Alice's station:\textcolor{black}{{}
We get ($\beta_{out}=\bar{g}\alpha$)} 
\begin{eqnarray}
F & = & \frac{2}{\sigma_{Q}}=\frac{2}{1+\bar{g}^{2}+E_{B|A}(\bar{g})}\,,\label{eq:fidelity_QAT-1-2-1}
\end{eqnarray}
where $E{}_{B|A}(\bar{g})$ is defined as the EPR steering parameter
of Eq.~(\ref{eq:eprsteercrit}) with $\mathbf{g}=(\bar{g},\bar{g})$.\textcolor{black}{{}
We have restricted to two-mode Gaussian resources with equal position
and momentum correlations \cite{zhangtele}, so that $|\langle X_{A},X_{B}\rangle|=|\langle P_{A},P_{B}\rangle|$,
$\sigma_{X}=\sigma_{P}$} and $\Delta(X_{B}-\bar{g}X_{A})=\Delta(P_{B}+\bar{g}P_{A})$.
This subclass, that we call $(X-P)$-balanced, includes EPR resources
such as the TMSS with phase-insensitive losses and noise. The special
BK case ($\bar{g}=1$) reduces to $F^{BK}=\frac{1}{(1+\Delta_{ent})}$,
where $\Delta_{ent}$ is the entanglement parameter of Eq.~(\ref{eq:duan-1-1-1}).
More generally, we see that the fidelity in Eq.~(\ref{eq:fidelity_QAT-1-2-1})
is sensitive to the steering parameter.

\textbf{\emph{Asymmetry and entanglement:}} It is known that Gaussian
steerable states not satisfying the Tan-Duan entanglement condition
$\Delta_{ent}<1$ exist, and can be created for example from a two-mode
squeezed state by adding asymmetric losses or thermal noise to each
of the EPR channels (Fig. 2) \cite{adesso,qhediscord}. These asymmetric
steerable states are \emph{not} useful for standard BK quantum teleportation,
which requires a fidelity of $F>1/2$ and hence a resource with $\Delta_{ent}<1$
\cite{zhangtele}.

\begin{figure}
\includegraphics[width=0.5\columnwidth]{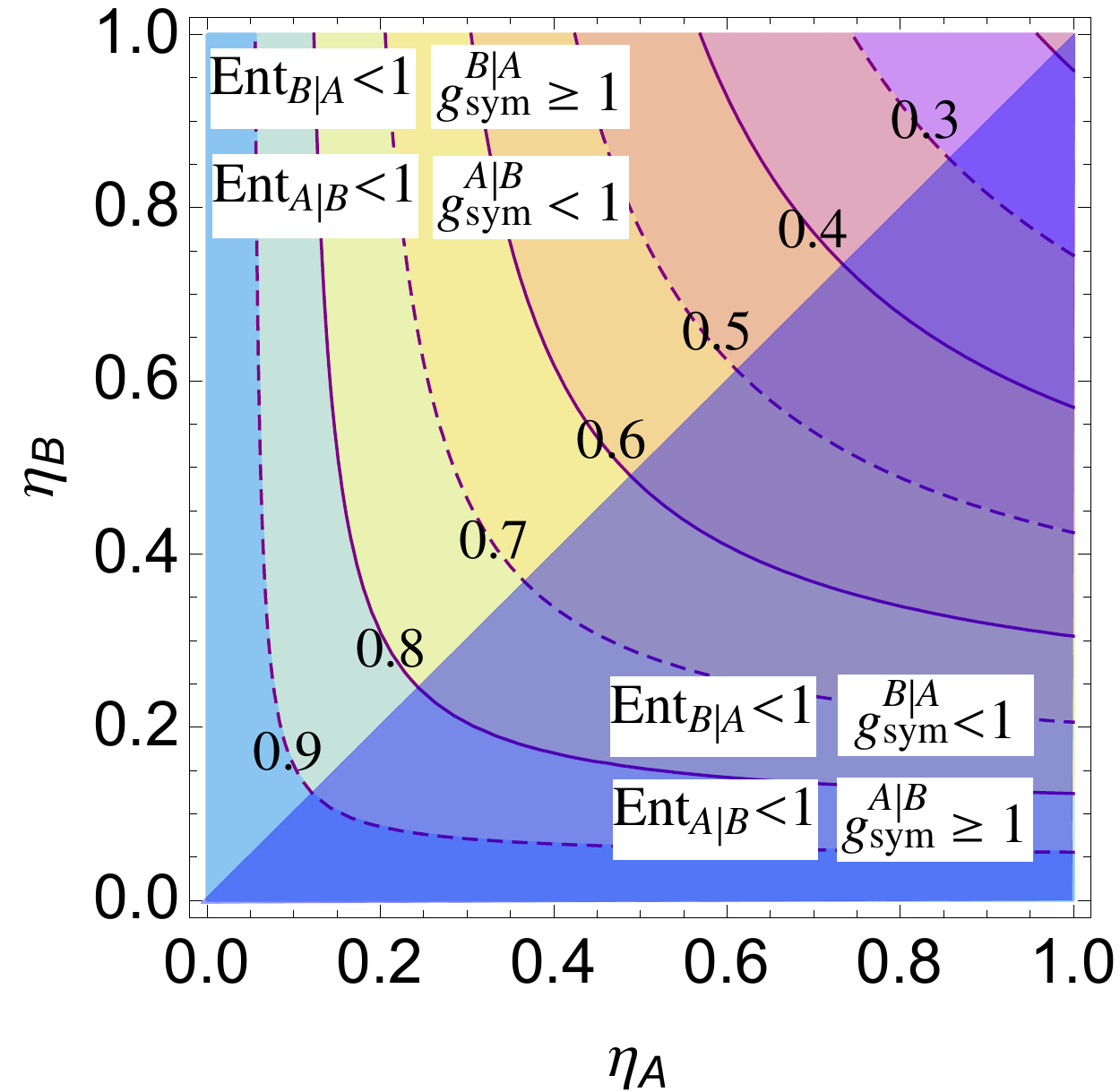}\includegraphics[width=0.5\columnwidth]{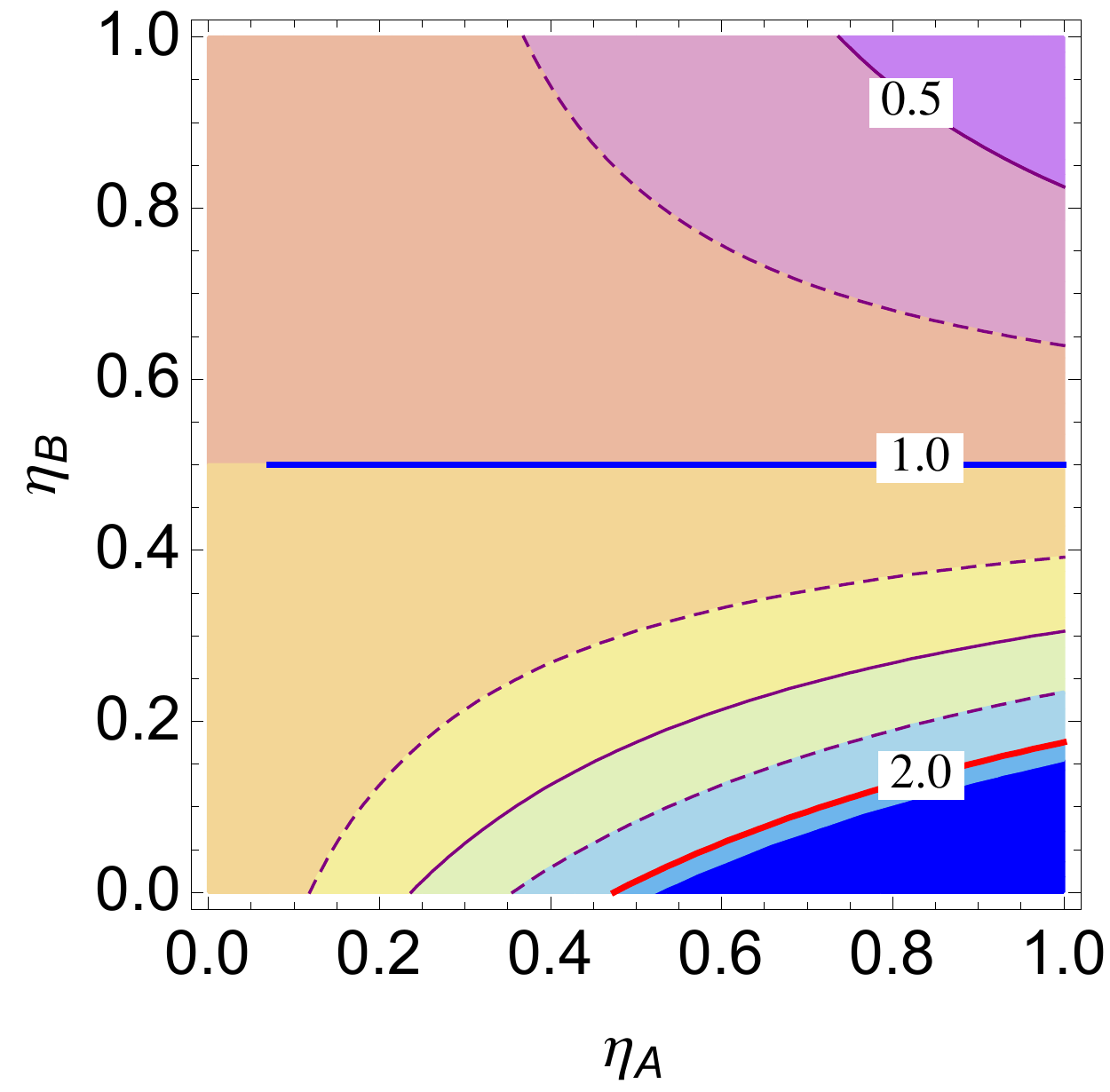}
\protect\protect\caption{\textcolor{black}{(Color online) }\emph{Steering for a two-mode squeezed
state with lossy channels:} Here $r=0.85$ and $\eta_{A/B}$ are the
channel efficiencies. \textbf{Left:}\emph{ }Contour lines show the
value of the entanglement parameter ${Ent}$. Higher loss on Alice's
EPR channel ($\eta_{A}<\eta_{B}$) implies $g_{sym}^{B|A}>1$ (note
$g_{sym}^{A|B}=1/g_{sym}^{B|A}$ and $\eta_{A}=\eta_{B}$ gives $g_{sym}^{B|A}=1$).
\textbf{Right: }Contour lines show the minimum value $E_{A|B}$ of
$E_{A|B}(g)$, found by selecting $g=c/m$. EPR steering of Alice
by Bob ($E_{A|B}<1$) is possible only when $\eta_{B}>1/2$.\label{fig:losses-2-2}\textcolor{red}{{} }}
\end{figure}

The first point we make is that the generalisation to non-unity classical
gains allows all the $(X-P)$-balanced Gaussian entangled states to
be useful for QT \cite{qhediscord}. We clarify as follows: The threshold
fidelity where one can rule out all classical measure-and-prepare
strategies as in Ref. \cite{ham} and hence claim QT is $F>\frac{1}{1+\bar{g}^{2}}$
(for $\bar{g}\geq1$) \cite{ham,g_ampli}. On examining (\ref{eq:fidelity_QAT-1-2-1}),
the condition on the resource to obtain QT reduces to: 
\begin{eqnarray}
{Ent}_{B|A}(\bar{g}) & = & \frac{\Delta(X_{B}-\bar{g}X_{A})\Delta(P_{B}+\bar{g}P_{A})}{(1+\bar{g}^{2})}<1.\label{eq:ent6}
\end{eqnarray}
The inequality ${Ent}_{B|A}(\bar{g})<1$, if satisfied, certifies
entanglement between any two modes $A$ and $B$ ($\bar{g}$ is any
real number)\textcolor{red}{{} }\cite{entcrit}. Further, the inequality
${Ent}_{B|A}(\bar{g})<1$ for an \emph{optimally selected $\bar{g}=g_{sym}^{B|A}$
}that minimises ${Ent}_{B|A}(\bar{g})$ \cite{qhediscord} is equivalent
to Simon's positive partial transpose condition for entanglement \cite{PPT-Peres},
which is necessary and sufficient for two-mode Gaussian states. The
parameter defined as ${Ent}\equiv{Ent}_{B|A}(g_{sym}^{B|A})$ is equal
to the lowest symplectic eigenvalue of the partial transpose $\nu$,
which determines the logarithmic negativity 
a measure of entanglement \cite{mari,adesso,adessotele,nu papers}.
Maximum entanglement corresponds to ${Ent}\rightarrow0$. Analysing
the result Eq.~(\ref{eq:ent6}), it is clear that for any Gaussian
entangled resource (within the ($X-P$)-balanced class) with $g_{sym}^{B|A}\geq1$,
we can quantum teleport $|\alpha\rangle\rightarrow|\bar{g}\alpha\rangle$
(from Alice to Bob) using the classical gain set at $\bar{g}=g_{sym}^{B|A}$
\cite{qhediscord}. When $g_{sym}^{B|A}<1$, QT is obtained by switching
the EPR channels $A$ and $B$.

The value $g_{sym}^{B|A}$ quantifies the asymmetry of the resource
and is calculated as $g_{sym}^{B|A}=x+\sqrt{x^{2}+1}$ where $x=(m-n)/2c$
and $n=\langle X_{A},X_{A}\rangle$, $m=\langle X_{B},X_{B}\rangle$,
$c=\langle X_{A},X_{B}\rangle=-\langle P_{A},P_{B}\rangle$. The coefficients
$n$, $m$, and $c$ fully define the covariance matrix of Gaussian
states in the $(X-P)$-balanced class. For a TMSS with losses at each
channel $A$ and $B$, so that $\eta_{A}$ and $\eta_{B}$ are the
respective efficiencies, the covariances become \textcolor{black}{$n=\eta_{A}\cosh\left(2r\right)+1-\eta_{A}$,
$m=\eta_{B}\cosh\left(2r\right)+1-\eta_{B}$, $c=\sqrt{\eta_{A}\eta_{B}}\sinh(2r)$}.
The entanglement and steering parameters for this resource are given
in Fig. 2. We next present a useful result that holds for all Gaussian
or non-Gaussian states with covariance matrix of the ($X-P$)-balanced
form.

\textbf{Result (1):} The amount of EPR entanglement is limited by
the asymmetry parameter, according to $\nu\equiv{Ent}\geq(g_{sym}^{2}-1)/(g_{sym}^{2}+1),$
Hence, two-way steering (corresponding to $\{E_{A|B},E_{B|A}\}<1)$
is certified if ${Ent}<1/(1+g_{sym}^{2})$. Two-way steerable states
are constrained to be reasonably symmetric, so that $g_{sym}<\sqrt{2}$,
and thus two-way steering is certified if 
\begin{equation}
\nu\equiv{Ent}<1/3.\label{eq:entonethrid}
\end{equation}
The condition is tight for ($X-P$)-balanced Gaussian states as shown
in Fig. 3a, and it agrees with that derived in \cite{adesso} in the
case of arbitrary two-mode Gaussian states. The proofs are given in
the Supplemental \cite{sup}.

\begin{figure}[tb]
\centering{}\includegraphics[bb=-1bp 4bp 543bp 494bp,clip,width=0.48\columnwidth]{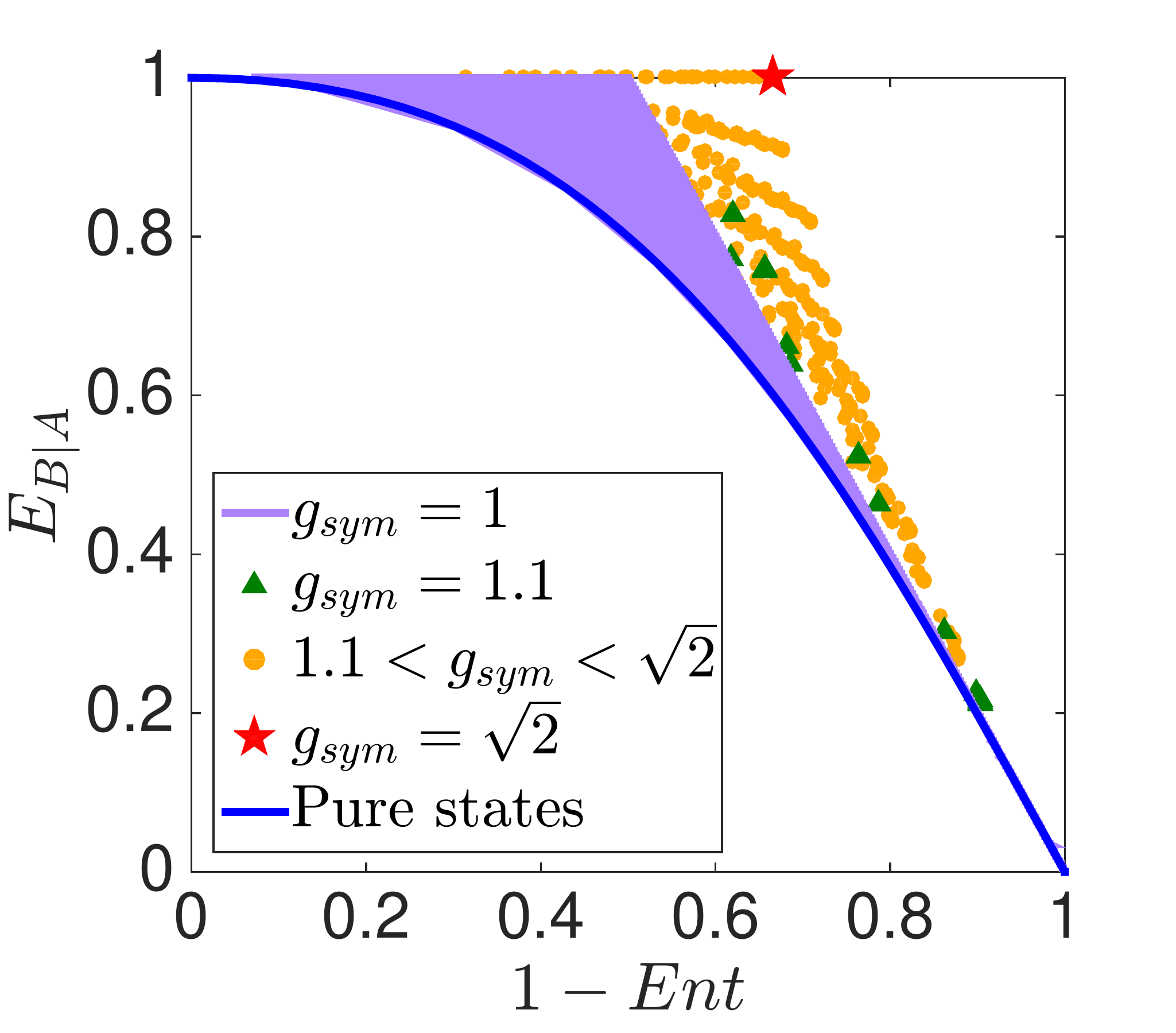}\includegraphics[width=0.5\columnwidth]{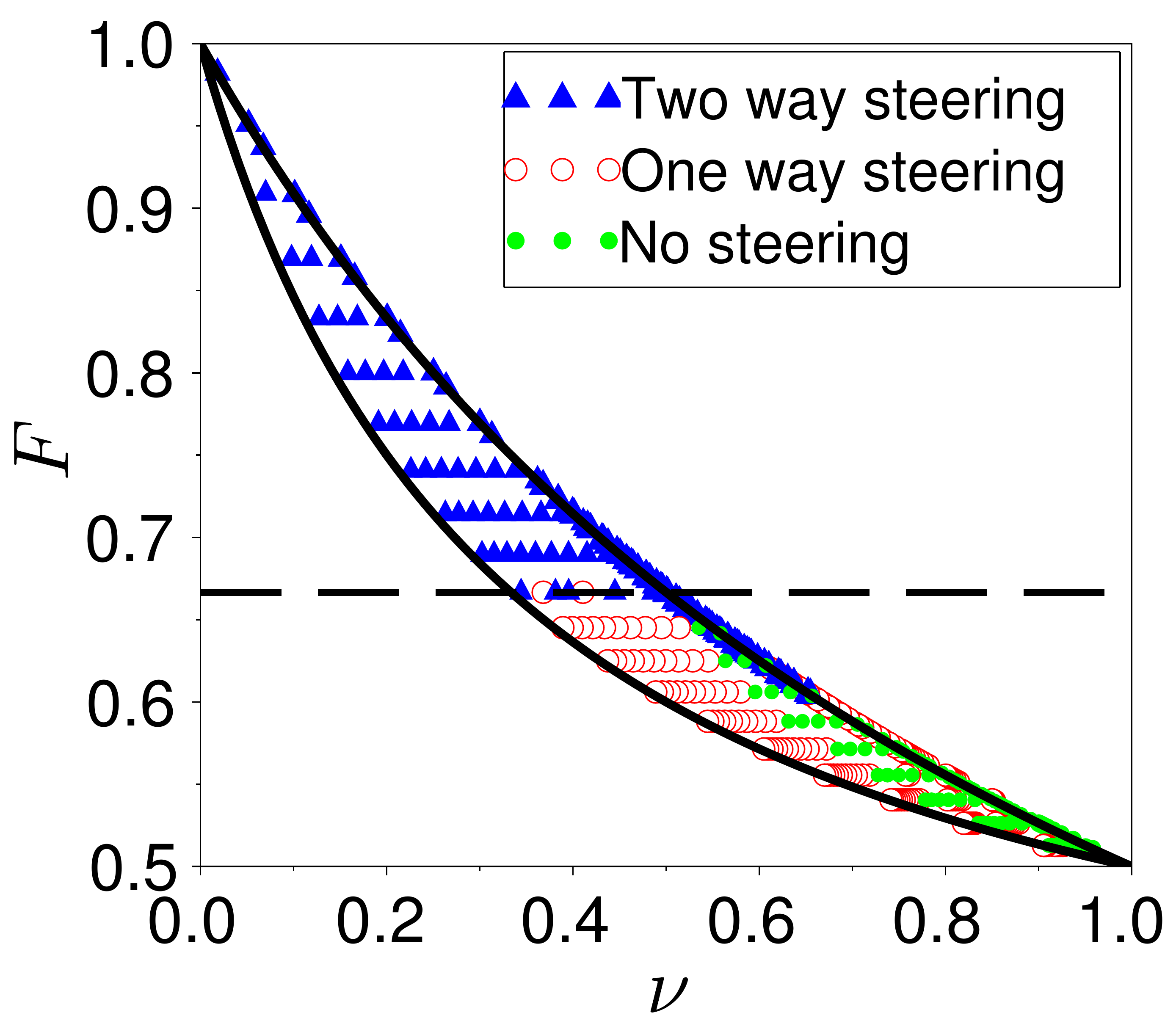}
\protect\protect\protect\caption{(Color online) \emph{Gaussian steerability, entanglement, asymmetry,
and teleportation fidelity.} \textbf{Left:} Gaussian steering of $B$
by $A$ occurs when \emph{$E_{B|A}<1$ }(maximum steering occurs as
$E_{B|A}\rightarrow0$). Entanglement occurs when ${Ent}<1$ (maximum
at ${Ent}\rightarrow0$). Here we use $g_{sym}^{B|A}>1$ for which
$E_{A|B}\leq E_{B|A}$. \label{fig:steeringvsdiscord-1} The purple
continuous region is for two-mode Gaussian symmetric states $g_{sym}=1$,
for which ${Ent}<1/2$ implies two-way steerability. The orange points
are for asymmetric Gaussian states, $g_{sym}>1$. The entanglement
condition ${Ent}<1/3$ implies two-way steerability for all $(X-P)$-balanced
states. \textbf{Right:} Optimal teleportation fidelity (using the
$BK$, $lsatt$ or $esa$ protocols, defined in the text) versus the
entanglement parameter $\nu={Ent}$ of the resource. Black-solid lines
denote the MV bounds. Steering properties of lossy TMSS resources
are plotted for all $\eta_{A}$, $\eta_{B}$ and $r\geq0.5$. Two-way
steering is required for secure teleportation of coherent states,
marked by $F>2/3$.}
\end{figure}

\textbf{\emph{Steering and secure teleportation:}} We can now address
the main question, namely what is the requirement on the resource
to achieve no-cloning ST. Restricting to lossy TMSS states, we can
see that a resource steerable $A$ by $B$ is required for ST. If
there is no steering of $A$ by $B$, the covariances imply that $\eta_{B}\leq1/2$
(Fig. 2). Such a field $B$ can be generated using a 50-50 beam splitter,
that produces a second field $B'$ satisfying $\langle X_{A},X_{B'}\rangle=\langle X_{A},X_{B}\rangle$,
$\langle P_{A},P_{B'}\rangle=\langle P_{A},P_{B}\rangle$, etc. This
implies that an observer Eve with access to $B'$ can generate from
the classical information (which is publicly accessible) the same
teleported state as can Bob, who has access only to field $B$ (see
Fig. 1). This argument, while restricted to the lossy TMSS resource,
is nonetheless quite powerful, applying to protocols with arbitrary
$\bar{g}$ and local operations at Bob's station, and (similar to
\cite{steer telequbit}) is not based on fidelity.

We now focus on the important case of conventional teleportation of
the coherent state $|\alpha\rangle\rightarrow|\alpha\rangle$, but
allowing for a broader set of Gaussian resources. Following GG, we
consider \emph{high fidelity} ST, where the no-cloning threshold for
security is established by $F>2/3$. GG revealed that for the BK protocol,
$F>2/3$ requires a resource with $\Delta_{ent}<1/2$. We now know
that this implies two-way steering of the resource \cite{qhediscord}.
Further, for symmetric resources, $g_{sym}=1$ and ${Ent}\equiv\Delta_{ent}$,
and we see from Result (1) and Fig. 3a that the GG condition $\Delta_{ent}<1/2$
is the \emph{tight} condition on the entanglement parameter $\nu={Ent}$,
for two-way steerability. This motivates us to generalise the GG result,
to include asymmetric Gaussian resources and protocols.

To teleport an unknown coherent state with optimal fidelity for a
given resource, local operations are needed at Alice and Bob's stations
(Fig. 1). The full optimisation for all protocols is difficult, but
Mari and Vitali (MV) optimised over all local Gaussian operations
to derive fidelity bounds for a given entanglement value $\nu={Ent}$,
given by \cite{mari} 
\begin{equation}
\mbox{\ensuremath{\frac{1+\nu}{1+3\nu}\leq F\leq\frac{1}{1+\nu}}}\,.\label{eq:MVbounds}
\end{equation}
The next result tells us that, once one allows asymmetric Gaussian
protocols, the set of resources enabling ST is expanded on those satisfying
the GG condition $\Delta_{ent}<1/2$.

\begin{figure}[t]
\centering{}\textcolor{black}{\includegraphics[width=0.5\columnwidth]{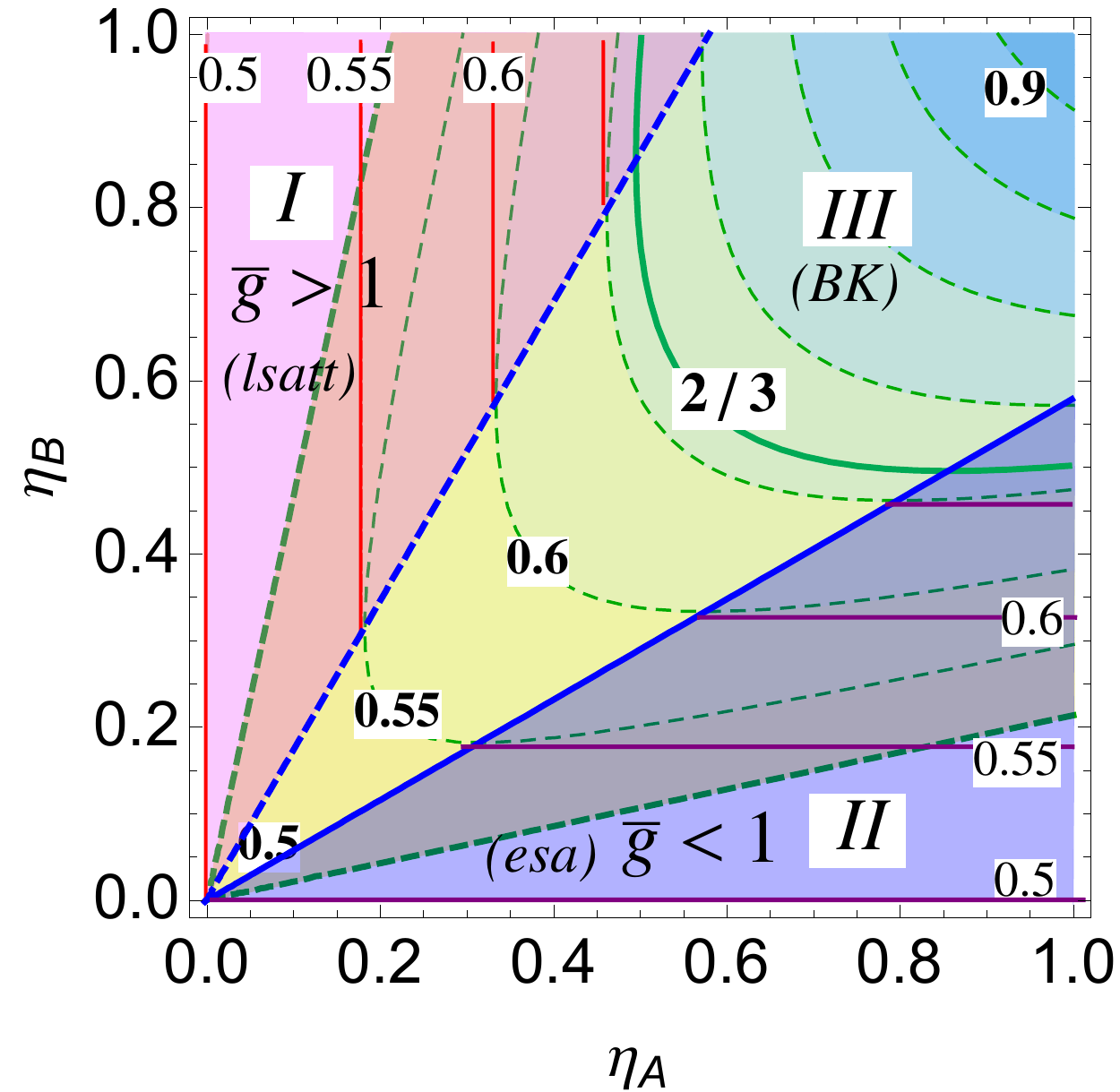}}\includegraphics[width=0.5\columnwidth]{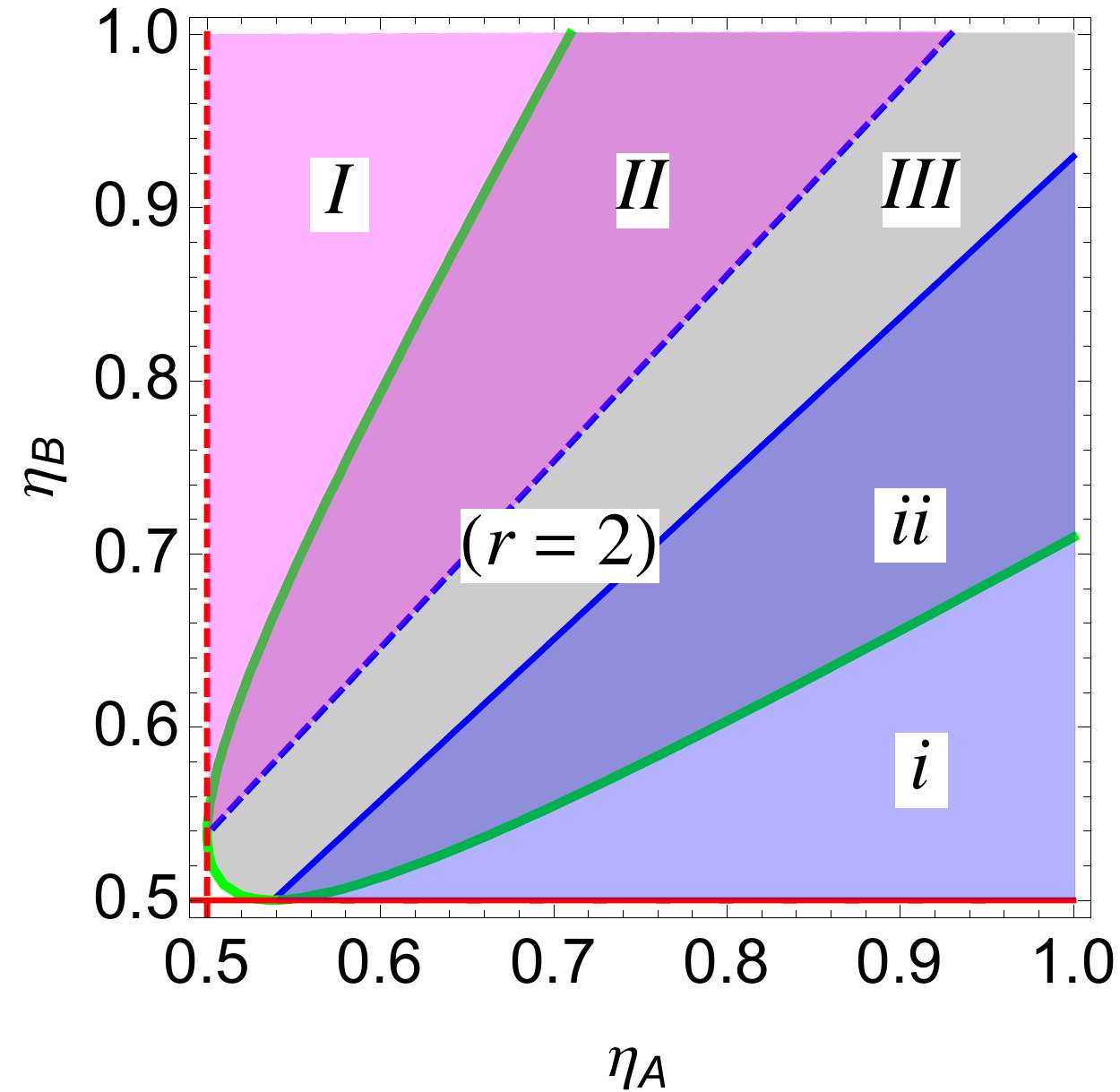}\protect\caption{(Color online) \emph{Optimising the fidelity for the teleportation
$|\alpha\rangle\rightarrow|\alpha\rangle$ from Alice to Bob, using
a resource with lossy channels.} \textbf{Left:} ($r=1.0$): Contours
show optimised fidelity values: We optimise via the lsatt protocol
(region $I$), or via the esa (region $II$), or via the BK protocol
(central coned region $III$). For higher $r>0.89$, ST ($F>2/3$)
is possible using the asymmetric protocol. \textbf{Right: }ST can
be achieved via the lsatt protocol (area $I+II$), or via the esa
(area $i+ii$). The green curve corresponds to $\Delta_{ent}<1/2$
so that regions $II$, $III$ and $ii$ give ST using the BK protocol.
Note that for area $III$, ST can be only achieved by the BK. Regions
$I$ and $i$ require asymmetric protocols for ST. \label{fig:atten_ampli_fidelities}}
\end{figure}

\textbf{Result (2):} All Gaussian entangled resources satisfying $\nu={Ent}<1/3$
are useful for ST of a coherent state. This entanglement threshold
is the same tight entanglement threshold to certify two-way steering,
Eq.~(\ref{eq:entonethrid}), see Fig. 3b. For symmetric resources
(where $g_{sym}=1$) the condition weakens, and all entangled resources
with ${Ent}<1/2$ are useful for ST.

\emph{Proof: }The MV bounds imply that for Gaussian states with $\nu={Ent}<1/3$,
an optimal protocol exists that will give a fidelity $F>2/3$. The
subsequent statements follow from Result (1) and the GG condition.

We remark that two-way steerable Gaussian entangled states exist that
satisfy $\nu={Ent}<1/3$ but do not satisfy the GG condition $\Delta_{ent}<1/2$.
For these states, the optimal protocol is not the BK one. It is hence
useful to understand how the optimal protocols can be carried out
in practice (Fig. 4). We present two simple protocols that are readily
achievable experimentally and that together with the BK protocol allow,
for any lossy TMSS resource, quantum teleportation with a fidelity
spanning the whole range within the MV bounds, Eq.~(\ref{eq:MVbounds}).
In fact, our study shows that for any such entangled Gaussian resource
with ${Ent}<1/3$, high fidelity ST can be carried out using one of
the three protocols that we call: lsatt, esa, or BK. The simplest
is \textbf{\emph{Late-stage attenuation (lsatt)}}: To teleport the
original state $|\alpha\rangle\rightarrow|\alpha\rangle$ when $\bar{g}>1$,
Bob locally attenuates his output field (Fig. 1). Bob may attenuate
using a beam splitter which yields a new output variance of $\sigma_{X}=\eta\sigma_{X}^{T}+1-\eta$
where $\eta=1/\bar{g}^{2}=g^{2}<1$, and $\sigma_{X}^{T}$ is the
variance for the original output. Then from Eq.~(\ref{eq:fidelity_QAT-1-2-1}),
the overall fidelity is $F_{A,B}^{lsatt}(g)=\frac{2}{3-g^{2}+E{}_{A|B}(g)}$
(using $\beta_{tele}=\alpha$). Standard QT with $F>1/2$ requires
a resource satisfying Eq.~(\ref{eq:ent6}), as for quantum teleamplification.
The fidelity is maximised for the choice of classical gain $\bar{g}_{opt}=(m-1)/c$,
and the overall optimal fidelity becomes $F_{A,B}^{lsatt}=\frac{2}{3+n-c^{2}/(m-1)}$,
provided $\bar{g}_{opt}>1$ ($m>c+1$) (see Fig. 4).

Alternatively, Alice may choose to\emph{ amplify} the input coherent
state at her station by a factor $g>1$, prior to a teleportation
protocol that uses a classical \emph{attenuation} factor $\bar{g}=1/g<1$.
We call this \textbf{\emph{Early-stage amplification (esa)}}. Suppose
Alice uses at her station a TMSS amplifier \cite{heraldednoiseless}.
Then the final amplified state at her station is given by a Gaussian
state with mean $g\alpha$ and variance $\sigma_{X/P,in}=2g^{2}-1$.
The final Gaussian output after teleportation to Bob has variance
$\sigma_{X}=\bar{g}^{2}\sigma_{X,in}+E{}_{B|A}(\bar{g})$. Substitution
into Eq.~(\ref{eq:fidelity_QAT-1-2-1}) reveals the fidelity for
the overall teleportation process to be $F_{A,B}^{esa}(\bar{g})=F_{B,A}^{lsatt}(\bar{g})$.
QT requires a resource satisfying the entanglement condition ${Ent}_{B|A}(\bar{g})<1$
with $\bar{g}<1$. Hence, for an entangled Gaussian resource (in the
$(X-P)$-balanced class) with $g_{sym}^{B|A}<1$, the esa protocol
with $\bar{g}=g_{sym}^{B|A}$ will give QT with $F>1/2$. The overall
fidelity $F_{A,B}^{esa}(\bar{g})$ is maximised for $\bar{g}_{opt}=c/(n-1)$
provided $\bar{g}_{opt}<1$ ($n>c+1$), in which case the fidelity
is given by $F_{A,B}^{esa}=F_{B,A}^{lsatt}$.\textcolor{black}{{} }

We finally give the connection with two-way steering.\\
 \textbf{ Result (3):} From the MV bounds, in order to achieve $F>2/3$,
the entanglement of the resource must satisfy ${Ent}<1/2$. Not all
resources with ${Ent}<1/2$ will allow $F>2/3$. Restricting to the
three protocols (lsatt, esa, BK), the requirement on the resource
to achieve the no-cloning fidelity $F>2/3$ is exactly for \emph{two-way
steering} (see Fig. 3b). The result is proved in the Supplemental
Materials \cite{sup}.

\textbf{\emph{Discussion:}} We conclude by suggesting a further application
of EPR steering to enhance the fidelity. The esa protocol relies on
pre-amplification of a coherent state by a factor $g>1$, which has
a limited maximum fidelity of $1/g^{2}$.\textbf{\emph{ }}Recent methods
propose heralding to overcome this limitation: heralded noiseless
amplification of $|\alpha\rangle$ to $|g\alpha\rangle$ potentially
allows fidelities approaching $1$ \cite{heraldednoiseless}. The
teleportation deamplification $|g\alpha\rangle\rightarrow|\alpha\rangle$
has fidelity $F=\frac{2}{1+\bar{g}^{2}+E{}_{B|A}(\bar{g})}$ given
by Eq.~(\ref{eq:fidelity_QAT-1-2-1}) but where $\bar{g}=1/g<1$.
We show in the Supplemental Material that for the TMSS resource with
the optimal choice of $\bar{g}$ ($\bar{g}_{opt}=\tanh(r)$), we get
$F\rightarrow1$. This is valid for all $r>0$, henceforth it does
not require significant entanglement of the teleportation resource.
The overall fidelity becomes limited by the fidelity of the heralded
amplification at Alice's site, suggesting a very promising experimental
route for achieving high teleportation fidelities. We note however
that there remains the requirement for EPR steerability of the resource,
which for low entanglement requires resources of sufficient purity
\cite{adesso}.

We thank the Australian Research Council for funding via Discovery
and DECRA grants. QYH thanks the support of the National Natural Science
Foundation of China under grants No.11274025, 61475006, and 11121091.
MR thanks J. Pryde and M. Hall for discussions. GA acknowledges discussions
with I. Kogias and thanks the European Research Council (ERC StG GQCOP
Grant No.~637352) for financial support.


\begin{thebibliography}{10}
\bibitem{notevariance} We consider two modes $A$ and $B$, with
boson annihilation operators $a$ and $b$, and number eigenstates
$|n\rangle_{A}$ and $|n\rangle_{B}$, respectively. Setting $\hbar=2$,
we define the quadratures as $X_{A}=a+a^{\dagger}$, $P_{A}=\left(a-a^{\dagger}\right)/i$,
$X_{B}=b+b^{\dagger}$ and $P_{B}=\left(b-b^{\dagger}\right)/i$,
and we use standard notation for the covariances $\langle X,Y\rangle=\langle XY\rangle-\langle X\rangle\langle Y\rangle$,
with the variance $\langle X,X\rangle\equiv\Delta X^{2}\equiv\sigma_{X}$
for the operator $X$.
\end{thebibliography}
\end{document}


\title{Supplemental Material}

\author{Q. Y. He$^{2}$, L. Rosales-Zárate{\normalsize{}{}{}$^{1}$}, G.
Adesso$^{3}$ and M. D. Reid{\normalsize{}{}{}$^{1}$} }

\affiliation{$^{1}$Centre for Quantum and Optical Science, Swinburne University
of Technology, Melbourne 3122 Australia}

\affiliation{$^{\text{2}}$State Key Laboratory of Mesoscopic Physics, School
of Physics, Peking University, Beijing 100871 China}

\affiliation{$^{\text{3}}$School of Mathematical Sciences, University of Nottingham,
Nottingham NG7 2RD, United Kingdom}

\maketitle

\section{Proofs}

\textbf{Result (1a):} The amount of EPR entanglement is limited by
the asymmetry parameter, according to $Ent\geq(g_{sym}^{2}-1)/(g_{sym}^{2}+1)$
where $g_{sym}=max(g_{sym}^{A|B},g_{sym}^{B|A})$.

\textbf{\emph{Proof:}} Without loss of generality, we will assume
$g_{sym}=g_{sym}^{B|A}=(m-n)/2c+\sqrt{[(m-n)/2c]^{2}+1}>1$, this
implies that $m\geq n$. Next, $Ent_{B|A}=[m-2cg_{sym}^{B|A}+n(g_{sym}^{B|A})^{2}]/[1+(g_{sym}^{B|A})^{2}]$,
using the definition of $g_{sym}$we can write $Ent_{B|A}$ and $(g_{sym}^{2}-1)/(g_{sym}^{2}+1)$
as: 
\begin{eqnarray*}
Ent_{B|A} & = & \frac{1}{2}\left(m+n-\sqrt{\left(m-n\right)^{2}+4c^{2}}\right),
\end{eqnarray*}

\[
\frac{g_{sym}^{2}-1}{g_{sym}^{2}+1}=\frac{\left(m-n\right)}{\sqrt{\left(m-n\right)^{2}+4c^{2}}}.
\]
We wish to prove that 
\[
\frac{1}{2}\left(m+n-\sqrt{\left(m-n\right)^{2}+4c^{2}}\right)=Ent_{B|A}\geq\frac{g_{sym}^{2}-1}{g_{sym}^{2}+1}=\frac{\left(m-n\right)}{\sqrt{\left(m-n\right)^{2}+4c^{2}}}.
\]
Expanding this expression we notice that it is equivalent to prove
that: 
\[
\left(m+n\right)^{2}c^{2}+\left(m-n\right)^{2}mn\geq\left(m-n\right)^{2}\left(1+m-n+2c^{2}\right)+2c^{2}\left(m-n+2c^{2}\right).
\]
From the PPT condition $nm-c^{2}+1-n-m>0$, we know that $nm>c^{2}-1+n+m$,
so we can write 
\begin{eqnarray*}
\left(m+n\right)^{2}c^{2}+\left(m-n\right)^{2}mn & \geq & \left(m+n\right)^{2}c^{2}+\left(m-n\right)^{2}(c^{2}-1+n+m).
\end{eqnarray*}
Next, we use that $-1+n+m\geq1+m-n$, this inequality holds since
we are considering $n\geq1$. Hence we can write: 
\begin{eqnarray*}
\left(m+n\right)^{2}c^{2}+\left(m-n\right)^{2}mn & \geq & \left(m+n\right)^{2}c^{2}+\left(m-n\right)^{2}(c^{2}-1+n+m)\\
 & \geq & \left(m+n\right)^{2}c^{2}+\left(m-n\right)^{2}(c^{2}+1+m-n)\\
 & \geq & c^{2}\left[\left(m+n\right)^{2}+\left(m-n\right)^{2}\right]+\left(m-n\right)^{2}(+1+m-n).
\end{eqnarray*}
So that in order to prove that $Ent\geq(g_{sym}^{2}-1)/(g_{sym}^{2}+1)$
we need to prove that: 
\begin{eqnarray*}
c^{2}\left[\left(m+n\right)^{2}+\left(m-n\right)^{2}\right]+\left(m-n\right)^{2}(1+m-n) & \geq & \left(m-n\right)^{2}\left(1+m-n+2c^{2}\right)+2c^{2}\left(m-n+2c^{2}\right).
\end{eqnarray*}
Let us prove this by contradiction, hence we suppose that 
\[
c^{2}\left[\left(m+n\right)^{2}+\left(m-n\right)^{2}\right]+\left(m-n\right)^{2}(1+m-n)\leq\left(m-n\right)^{2}\left(1+m-n+2c^{2}\right)+2c^{2}\left(m-n+2c^{2}\right).
\]
This implies that:

\begin{eqnarray*}
 &  & c^{2}\left[\left(m+n\right)^{2}+\left(m-n\right)^{2}\right]+\left(m-n\right)^{2}(1+m-n)\leq\left(m-n\right)^{2}\left(1+m-n+2c^{2}\right)+2c^{2}\left(m-n+2c^{2}\right)\\
\iff &  & c^{2}\left[\left(m+n\right)^{2}+\left(m-n\right)^{2}\right]\leq2c^{2}\left(m-n\right)^{2}+2c^{2}\left(m-n+2c^{2}\right)\\
\iff &  & \left(m+n\right)^{2}\leq\left(m-n\right)^{2}+2\left(m-n+2c^{2}\right)\\
\iff &  & m^{2}+2mn+n^{2}\leq m^{2}-2mn+n^{2}+2m-2n+4c^{2}\\
\iff &  & 4mn\leq2m-2n+4c^{2}\\
\iff &  & 2\left(mn-c^{2}\right)\leq m-n.
\end{eqnarray*}
But this is a contradiction since $2(nm-c^{2})\geq m-n$, this is
from PPT $nm-c^{2}>-1+n+m$, so that $2(nm-c^{2})>-2+2n+2m\geq2m\geq m\geq m-n$
since $n-1\geq0$. Therefore $Ent\geq(g_{sym}^{2}-1)/(g_{sym}^{2}+1)$
as required.

\emph{Proof for Gaussian case:} Note that the Result (1a) is easily
derived in the Gaussian case, since the fidelity of QAT is limited
to $1/\bar{g}^{2}$ for any $\bar{g}\geq1$. Hence, from the expression
for the fidelity (Eq. (3)) given in the main paper, we see that $Ent_{B|A}(\bar{g})\geq(\bar{g}^{2}-1)/(\bar{g}^{2}+1)$.
Putting $\bar{g}=g_{sym}$, the result follows. The above proof is
general for two-mode states.

\textbf{Result (1b): }Hence, two-way steering is certified if 
\[
Ent<1/(1+g_{sym}^{2}).
\]
Two-way steerable states require a minimum symmetry, $g_{sym}<\sqrt{2}$
and thus, \emph{two-way steering} is certified if 
\[
Ent<1/3.
\]

\textbf{\emph{Proof:}}\textbf{ }Let us suppose without loss of generality
that $g_{sym}=g_{sym}^{B|A}\geq1$, then $Ent=Ent_{B|A}(g_{sym})$.
Steering of $B$ by $A$ is realised if $E_{B|A}(g_{sym})<1$ or equivalently
$Ent=Ent_{B|A}(g_{sym})<1/(1+g_{sym}^{2})$. We note that if $E_{B|A}(g_{sym})<1$,
then $E_{A|B}(1/g_{sym})<1/g_{sym}^{2}$ which for $g_{sym}\geq1$
implies steering of $A$ by $B$, and thus two-way steering. From
Result (1), for $Ent<1/(1+g_{sym}^{2})$ to be possible, we need $1\leq g_{sym}<\sqrt{2}$.
If $g_{sym}=1$, then $Ent<0.5$ is sufficient for two-way steering.
If $1\leq g_{sym}<\sqrt{2}$ then $Ent<1/3$ is sufficient for two-way
steering. If $g_{sym}\geq\sqrt{2}$, then from Result (1), it is impossible
to obtain $Ent<1/3$. Thus $Ent<1/3$ is a sufficient criterion for
two-way steering, for any value of $g_{sym}$.

\textbf{Result (3):} From the MV bounds, in order to achieve $F>2/3$,
the entanglement of the resource must satisfy $Ent<1/2$. Not all
resources with $Ent<1/2$ will allow $F>2/3$. Restricting to the
three protocols (lsatt, esa, BK), the requirement on the resource
to achieve the no-cloning fidelity $F>2/3$ is necessarily \emph{two-way
steerable}.

\textbf{\emph{Proof:}}\emph{ }For the lsatt protocol, $F>2/3$ leads
to a steering condition $E{}_{A|B}(g)<g^{2}<1$. On dividing through
by $g^{2}$, we see that this condition also implies steering of $B$
by $A$ i.e. $E{}_{B|A}(\bar{g})<1,$ where $\bar{g}=1/g$. The result
for the esa protocol is similar. For the BK protocol $\bar{g}=1$,
the requirement $F>2/3$ imposes the condition $\Delta_{ent}<0.5$
derived by Grosshans and Grangier \cite{Grangier clone}, and therefore
using results proved in Ref. \cite{qhediscord}, also the condition
that the resource be two-way steerable.

\section{Fidelity for quantum tele-amplification (QAT)}

The strategy outlined in the main paper of selecting $\bar{g}=g_{sym}$
will optimise the fidelity of the amplified teleportation protocol
\textit{relative to }the quantum benchmark $1/(1+\bar{g}^{2})$ as
given in the main paper. Immediately, we see that for all \textit{symmetric}
resources, the BK protocol is optimal in giving the maximum relative
fidelity for the amplified teleportation scheme. For \emph{asymmetric
}resources however, the BK protocol is \emph{not }optimal.

\begin{figure}[H]
\begin{centering}
\includegraphics[width=0.6\columnwidth]{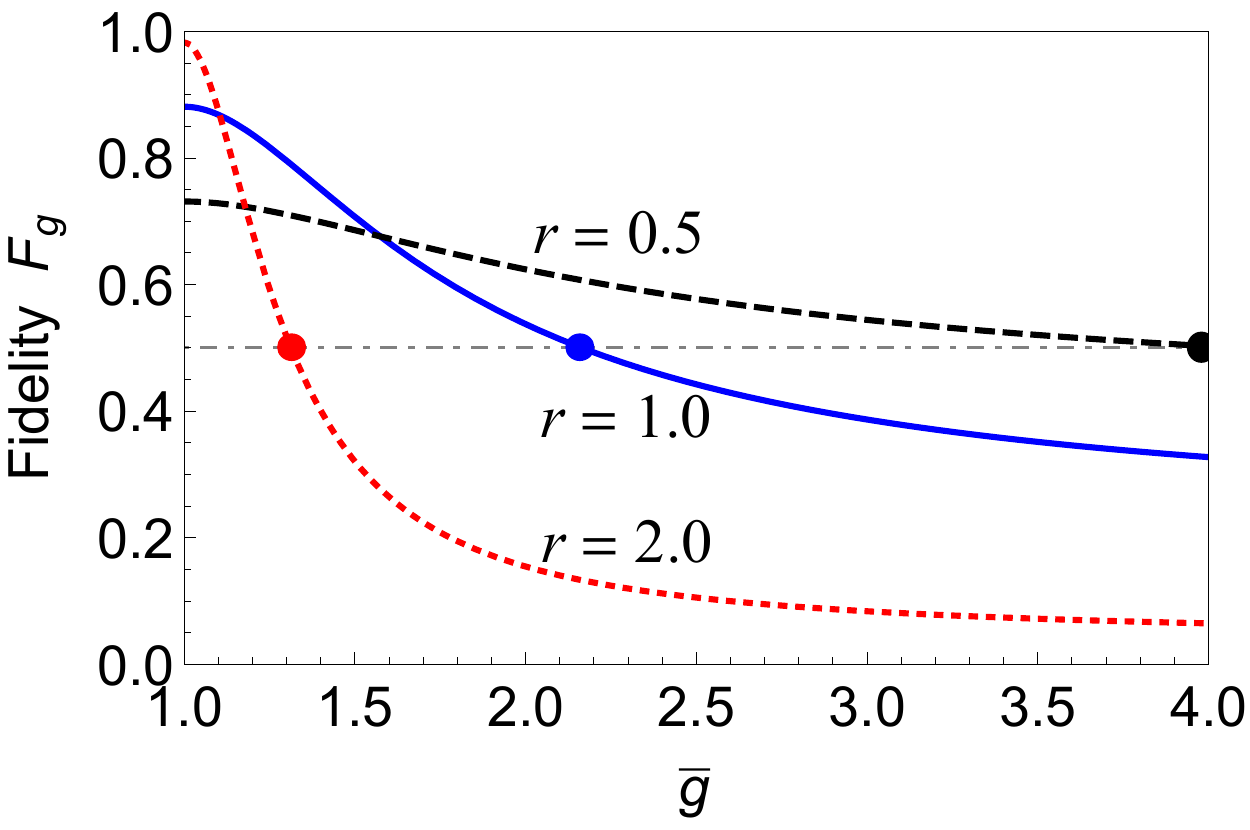} 
\par\end{centering}

\protect\protect\protect\caption{The fidelity $F_{g}\equiv\frac{F}{2/(1+\bar{g}^{2})}$ for amplified
teleportation $|\alpha\rangle\rightarrow|\bar{g}\alpha\rangle$ \textit{relative
to} the quantum benchmark, using a two-mode squeezed state resource.
Here, $\bar{g}$ is the classical gain and $r$ is the squeezing parameter
that determines the amount of entanglement of the resource. QT is
obtained for $F_{g}>0.5$. We see that for quantum tele-amplification
$\bar{g}>1$, the fidelity is optimised at a lower entanglement (i.e.
lower squeezing parameter $r$) value of the resource. \label{fig:relative_fidelity}}
\end{figure}

\textcolor{black}{Figure }\ref{fig:relative_fidelity}\textcolor{black}{{}
plots the relative fidelity $F_{g}$ for the two-mode squeezed state
(TMSS). In fact, the highly entangled TMSS cannot give QAT for high
gains $\bar{g}$: the TMSS with lower $r$ values is better for higher
gain $\bar{g}$, as is the resource created by adding asymmetric loss.}

We note that for all symmetric pure states, $g_{sym}=1$ and the BK
protocol gives the optimal relative fidelity. The TMSS is pure and
symmetric and will enable quantum teleportation (QT) using the BK
protocol, because there is symmetric entanglement that satisfies the
Duan entanglement condition for all $r$. This is evident from Fig.
\ref{fig:relative_fidelity}. Interestingly, we see that the TMSS
allows QT for gains up to $\coth(r/2)$ (three circles on the curves).
It is interesting that at higher values of $\bar{g}$, QT is obtained
by \emph{reducing} the entanglement (decreasing the value of $r$)
of the TMSS resource.

\begin{figure}
\begin{centering}
\includegraphics[width=0.6\columnwidth]{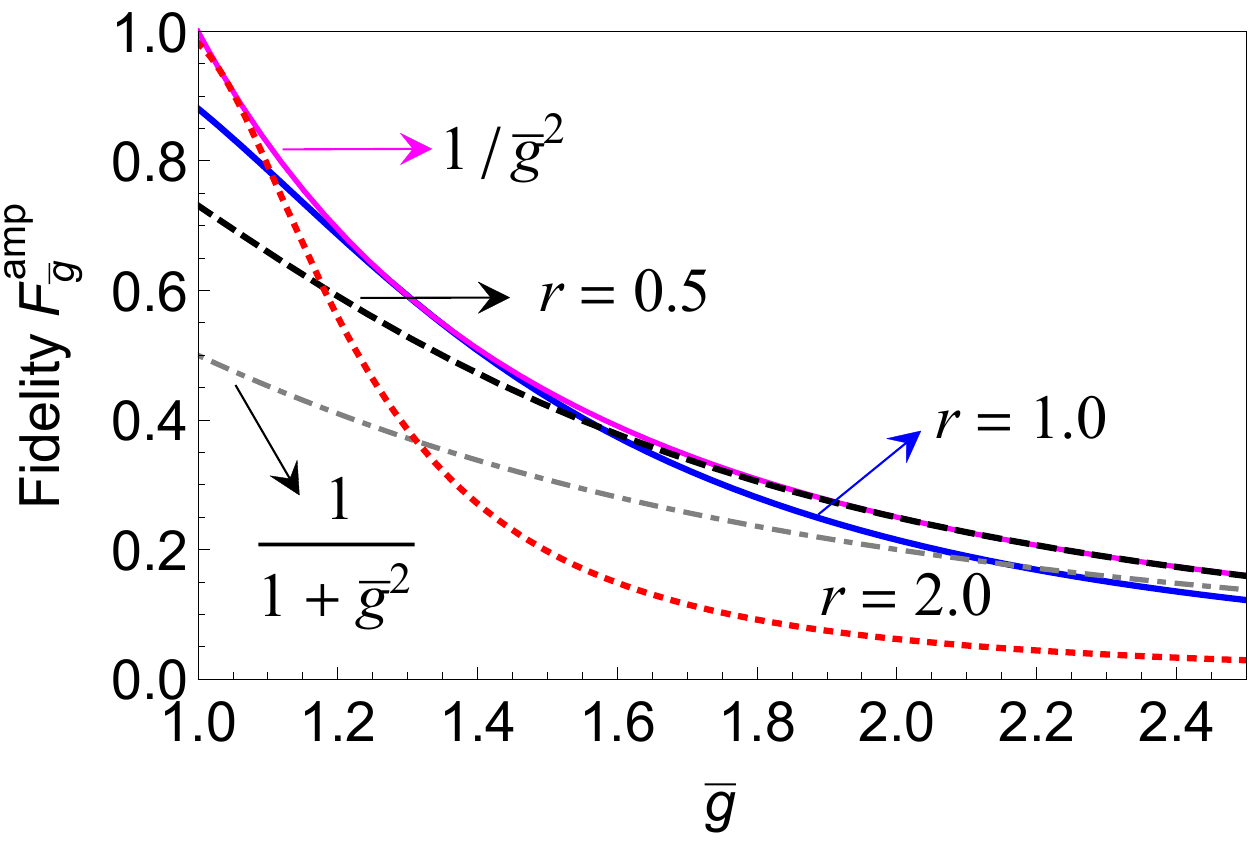} 
\par\end{centering}

\protect\protect\protect\caption{(Color online) The absolute fidelity for amplified teleportation $|\alpha\rangle\rightarrow|\bar{g}\alpha\rangle$
(Eq. (3) of main text) versus $\bar{g}$ using a two-mode squeezed
state resource. Here, $\bar{g}$ is the classical gain and $r$ is
the squeezing parameter that determines the amount of entanglement
of the resource. The threshold for QT is $F_{\bar{g}}^{amp}>\frac{1}{1+\bar{g}^{2}}$.
Again, we see that for larger amplification $\bar{g}$ the fidelity
is increased by decreasing the squeezing parameter $r$ of the TMSS
resource. \label{fig:absolute_fidelity}}
\end{figure}

We now consider how to optimise for the maximum \emph{absolute} fidelity
$F\equiv F_{\bar{g}}^{amp}$ of the amplified teleportation QAT process:
$|\alpha\rangle\rightarrow|\bar{g}\alpha\rangle$ (Alice to Bob),
where $\bar{g}>1$. \textcolor{black}{Thus we ask what is the maximum
}\textcolor{black}{\emph{absolute}}\textcolor{black}{{} value of
fidelity $F_{\bar{g}}^{amp}=2/(1+\bar{g}^{2}+E_{B|A}(\bar{g}))$ (as
given by Eq. (3) of the main text), for a given EPR resource. We need
to minimise $1+\bar{g}^{2}+E{}_{B|A}(\bar{g})$ which leads to the
condition $\bar{g}=\bar{g}_{opt}$ where} 
\begin{equation}
\bar{g}_{opt}=\frac{\langle X_{A},X_{B}\rangle}{1+\langle X_{A},X_{A}\rangle}\label{eq:gbar-1}
\end{equation}
as the optimal gain, provided $\bar{g}_{opt}>1$. In this case, the
maximum fidelity is 
\begin{equation}
F_{\bar{g}_{opt}}^{amp}=\frac{2(1+\langle X_{A},X_{A}\rangle)}{(\langle X_{B},X_{B}\rangle+1\text{)}(1+\langle X_{A},X_{A}\rangle)-\langle X_{A},X_{B}\rangle^{2}}.\label{eq:maxfid-1}
\end{equation}
For TMSS case, we have $\langle X_{A},X_{B}\rangle/(1+\langle X_{A},X_{A}\rangle)=\sinh(2r)/[1+\cosh(2r)]=2\sinh(r)\cosh(r)/[2\cosh^{2}(r)]=\sinh(r)/\cosh(r)<1$
because \textcolor{black}{$n=\cosh\left(2r\right)$, $m=\cosh\left(2r\right)$,
and $c=\sinh(2r)$}. If the value $\bar{g}_{opt}\leq1$, then the
optimal gain is $\bar{g}=1$, and the maximum fidelity for the TMSS
is given by the BK protocol. We note that the maximum achievable fidelity
for \emph{any} amplifier process $|\alpha\rangle\rightarrow|\bar{g}\alpha\rangle$
is $1/\bar{g}^{2}$, as shown by the magenta curve in Fig. \ref{fig:absolute_fidelity}.

Since for the pure TMSS, we cannot select the optimal $\overline{g}_{opt}$
and the corresponding maximum fidelity given by Eq. (\ref{eq:maxfid-1}).
However, we plot the \emph{absolute} fidelity $F_{\bar{g}}^{amp}$
versus the amplification gain $\overline{g}\geq1$, in Fig. \ref{fig:absolute_fidelity}.
We find as expected that the maximum absolute fidelity is for\textcolor{blue}{{}
}\textcolor{black}{$\bar{g}=1$, corresponding to the BK protocol.}

\section{The fidelity for attenuation as part of a preamplification protocol}

\begin{figure}
\begin{centering}
\includegraphics[width=0.6\columnwidth]{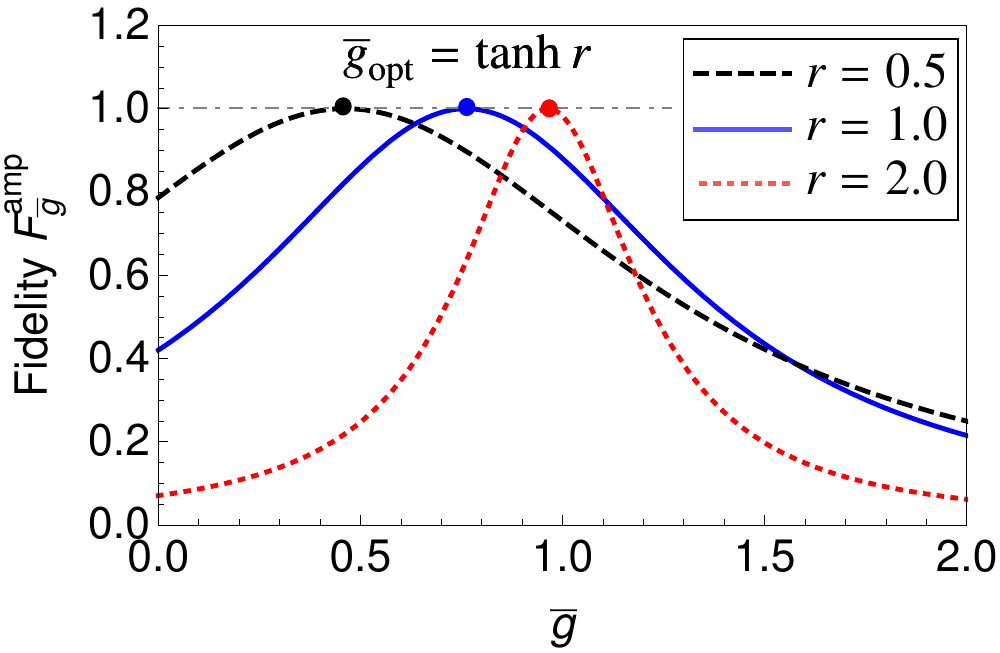} 
\par\end{centering}

\protect\protect\protect\caption{(Color online) \textcolor{black}{The maximum }\textcolor{black}{\emph{absolute}}\textcolor{black}{{}
value of fidelity $F_{\bar{g}}$ possible for the attenuation teleportation
process $|\alpha\rangle\rightarrow|\bar{g}\alpha\rangle$} where $\bar{g}<1$
is $F_{\bar{g}_{opt}}=1$ given by TMSS with the condition $\bar{g}_{opt}=\tanh(r)$.\label{fig:absolute_fidelity-1}}
\end{figure}

We can then ask what is the maximum \emph{absolute} value of fidelity
possible for the process $|\alpha\rangle\rightarrow|\bar{g}\alpha\rangle$
shown by $F_{\bar{g}}^{amp}=2/(1+\bar{g}^{2}+EPR_{B|A}(\bar{g}))$,
where we consider attenuation $\bar{g}<1$, for a given EPR resource.
We need to minimise $1+\bar{g}^{2}+EPR_{B|A}(\bar{g})$ which leads
to the condition $\bar{g}=\bar{g}_{opt}$ where 
\begin{equation}
\bar{g}_{opt}=\frac{\langle X_{A},X_{B}\rangle}{1+\langle X_{A},X_{A}\rangle}\label{eq:gbar-1-1}
\end{equation}
as being the optimal gain, provided $\bar{g}_{opt}<1$. In this case,
the maximum fidelity is 
\begin{equation}
F_{\bar{g}_{opt}}=\frac{2(1+\langle X_{A},X_{A}\rangle)}{(\langle X_{B},X_{B}\rangle+1\text{)}(1+\langle X_{A},X_{A}\rangle)-\langle X_{A},X_{B}\rangle^{2}}.\label{eq:maxfid-1-1}
\end{equation}
For the TMSS, we have $\langle X_{A},X_{B}\rangle/(1+\langle X_{A},X_{A}\rangle)=\sinh(2r)/[1+\cosh(2r)]=2\sinh(r)\cosh(r)/[2\cosh^{2}(r)]=\sinh(r)/\cosh(r)<1$
because \textcolor{black}{$n=\cosh\left(2r\right)$, and $m=\cosh\left(2r\right)$,
$c=\sinh(2r)$}. Therefore, the maximum fidelity is given by 
\begin{equation}
F_{\bar{g}_{opt}}=\frac{2[1+\cosh(2r)]}{[\cosh(2r)+1]^{2}-\sinh^{2}(2r)}=\frac{2[1+\cosh(2r)]}{2[1+\cosh(2r)]}=1,\label{eq:maxfid-1-1-1}
\end{equation}
as shown by three circles on curves in Fig. \ref{fig:absolute_fidelity-1}.

Thus, to implement the strategy, one would want to preamplify $|\alpha\rangle\rightarrow|g\alpha\rangle$
($g=1/\bar{g}>1$) using a heralded noiseless amplification technique.
The following stage is the attenuation which gives maximum fidelity
of one, for arbitrary $r$. Thus the fidelity will largely be determined
by the fidelity of the heralded preamplification. This will be limited
in practice by the $g$ value (and the $\alpha$ value) given current
experimental limitations. However, taking $g\sim2$, we would need
$\bar{g}=0.5=\tanh(r)$ which implies $r=0.55$. This is a quite accessible
squeezing parameter.

\section{Fidelity using BK and LSATT protocols }

\begin{figure}
\begin{centering}
\includegraphics[width=0.5\columnwidth]{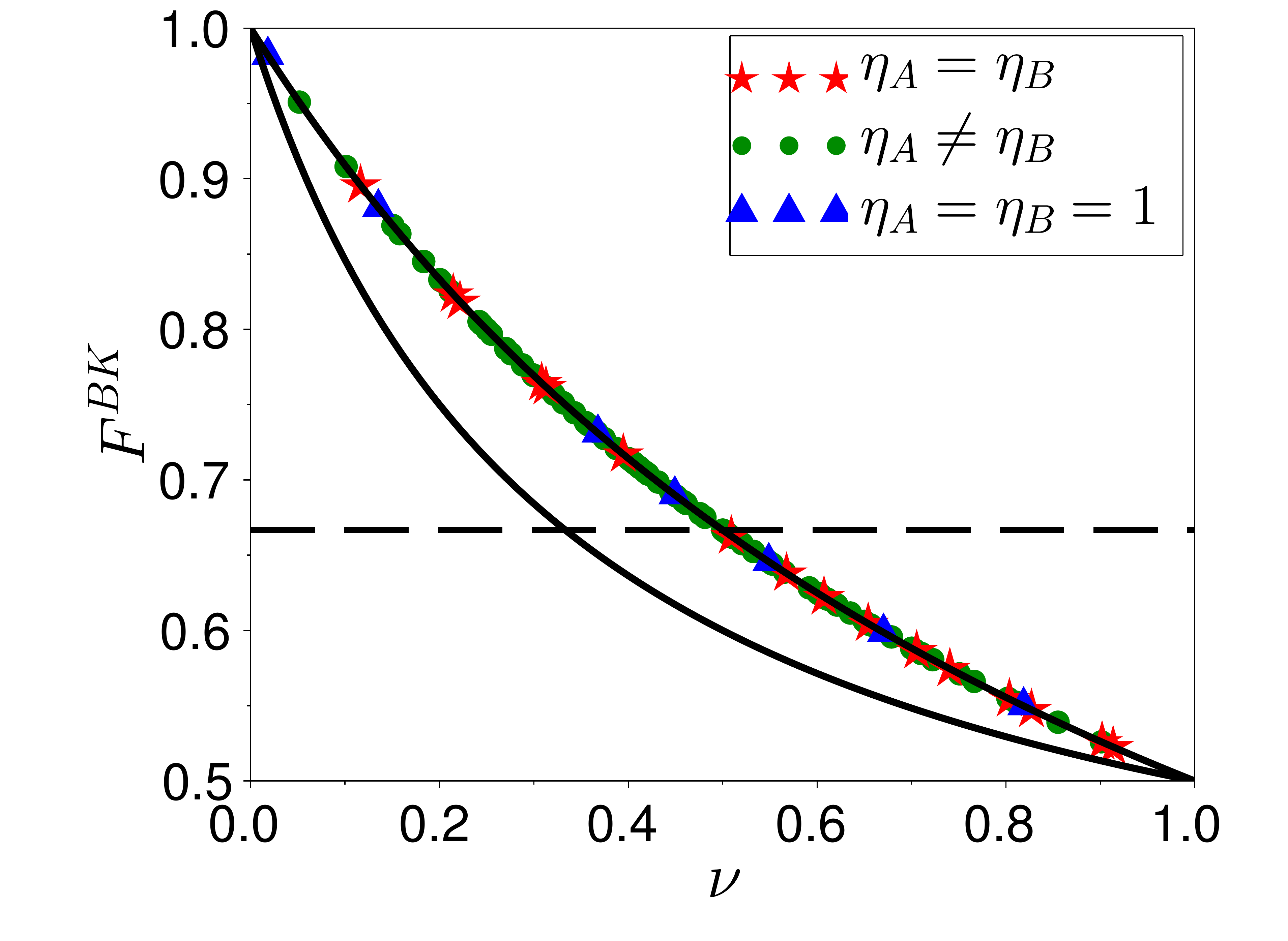}\includegraphics[width=0.5\columnwidth]{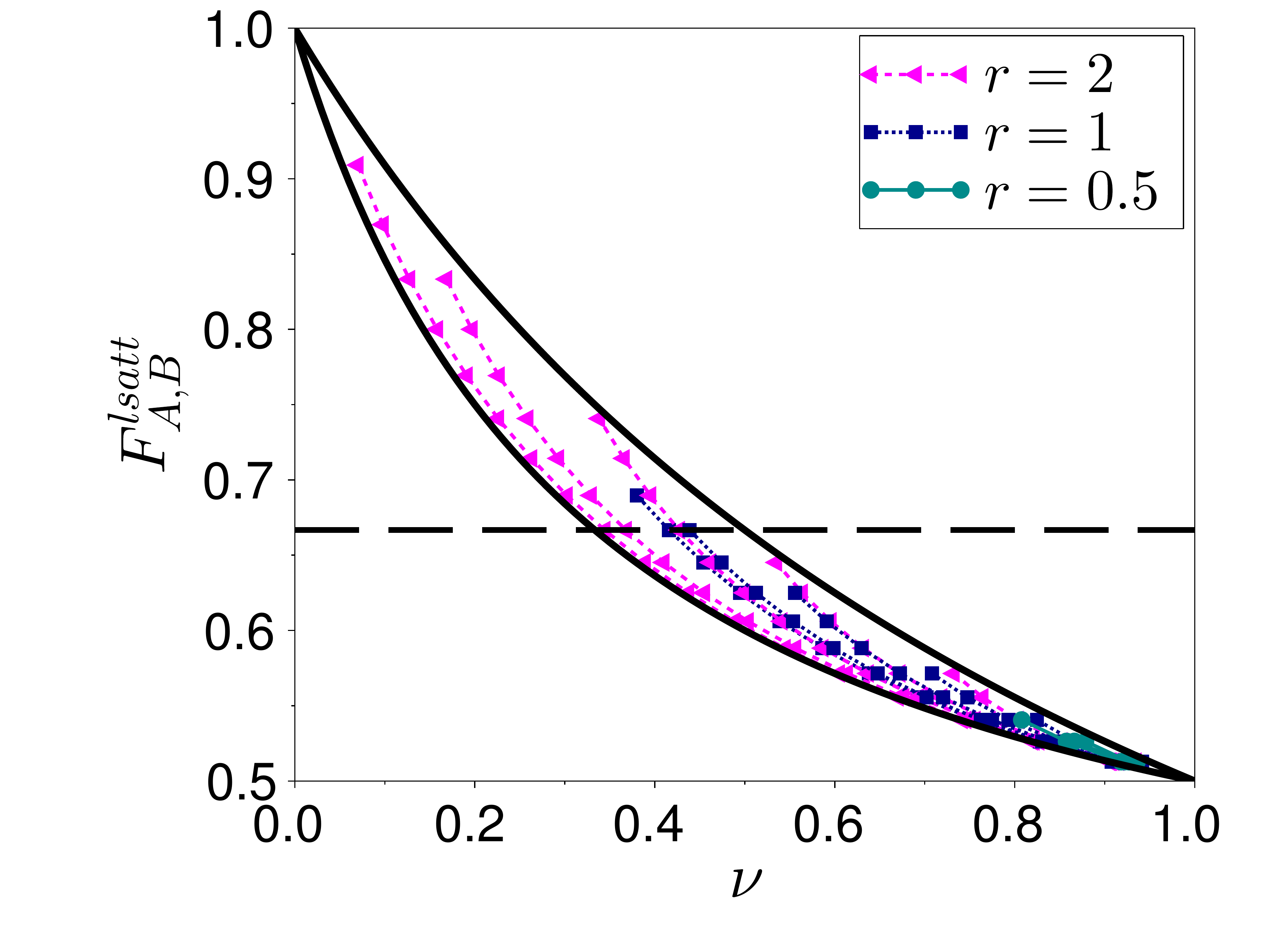} 
\par\end{centering}

\protect\protect\caption{(Color online) Optimal teleportation fidelity versus entanglement
$\nu$ of the resource. \textbf{Left:} with BK protocol, the blue
triangles correspond to pure TMSS resource. Red stars correspond to
mixed states with $\eta_{A}=\eta_{B}$, while green circles correspond
to mixed states with $\eta_{A}\protect\neq\eta_{B}$. Here we have
used $r=2$ and $r=1$. Black-solid lines denote MV bounds. \textbf{Right:}
with lsatt protocol, the curves with pink triangles, blue squares,
and dark cyan points are asymmetric mixed states with $r=2$, $r=1$
and $r=0.5$, respectively. The lines correspond to fixed values of
$r$, we have used the following values of $\eta_{A}=1,\:0.9,\:0.5,\:0.3,\:,0.1$
and $\eta_{B}=0.1,\:0.2,\:\ldots,\:1$ respectively. Black-solid lines
denote MV bounds. \label{fig:Opt_fidelity} }
\end{figure}

It is useful to understand how the optimal protocols can be carried
out. Figure \ref{fig:Opt_fidelity} presents the optimal teleportation
fidelity carried out with BK (left) and lsatt (right) protocols for
TMSS resource with losses. Here we have considered the expressions
for the fidelity for each protocol which are given in the main paper.
These expressions are:
\begin{eqnarray*}
F^{BK} & = & \frac{1}{1+\Delta_{ent}},\\
F_{A,B}^{lsatt} & = & \frac{2}{3-g_{opt}^{2}+E_{A|B}(g_{opt})}.
\end{eqnarray*}
 Using BK protocol, secure teleportation ($F>2/3$) is achievable
when $\nu\equiv\Delta_{ent}<0.5$ with symmetric and asymmetric resources.
 For both protocols, secure teleportation is harder to achieve when
$r\rightarrow0$. We require $r>0.347$ for $\eta_{A}=\eta_{B}=1$
to achieve ST for BK which is a lower bound than for lsatt (refer
Fig. 4 of the main text). For both protocols the values of optimal
fidelity lie in the MV bounds, indicated by the black solid lines.

